\documentclass[a4paper,12pt]{article}
\usepackage[pctex32]{graphics}

\begin{document}
\topmargin 0pt \oddsidemargin 0mm
\newcommand{\beq}{\begin{equation}}
\newcommand{\eeq}{\end{equation}}
\newcommand{\beqa}{\begin{eqnarray}}
\newcommand{\eeqa}{\end{eqnarray}}
\newcommand{\fr}{\frac}
\renewcommand{\thefootnote}{\fnsymbol{footnote}}
\newcommand{\rf}[1]{\ref{#1}}
\begin{titlepage}
\begin{flushright}
INJE-TP-05-04\\
hep-th/0506031
\end{flushright}

\vspace{5mm}
\begin{center}
{\Large \bf Unitarity issue in BTZ black holes} \vspace{12mm}

{\large  Y. S. Myung$^{1,2}$\footnote{e-mail
 address: ysmyung@physics.inje.ac.kr} and H. W. Lee$^1$\footnote{Email-address :
hwlee@inje.ac.kr} }
 \\
\vspace{10mm} {\em  $^{1}$Institute of Mathematical Science and
School of Computer Aided School \\ Inje University,
Gimhae 621-749, Korea \\
$^{2}$Institute of Theoretical Science, University of Oregon,
Eugene, OR 97403-5203, USA}
\end{center}

\begin{abstract}
We study the wave equation for  a  massive scalar  in
three-dimensional AdS-black hole spacetimes to understand the
unitarity issues in a semiclassical way. Here we introduce four
interesting spacetimes: the non-rotating BTZ black hole (NBTZ),
 pure AdS spacetime (PADS), massless BTZ black
hole (MBTZ), and  extremal BTZ black hole (EBTZ). Our method is
based on the potential analysis and solving the wave equation to
find  the condition for the frequency $\omega$ exactly.   In the
NBTZ case, one finds the quasinormal (complex and discrete) modes
which signals for a non-unitary evolution. Real and discrete modes
are found for the PADS case, which means that it is unitary
obviously. On the other hand, we find real and continuous modes
for the two extremal black holes of MBTZ and EBTZ. It suggests
 that these could be candidates for the unitary system.

\end{abstract}
\end{titlepage}

\newpage
\setcounter{page}{2}

 \section{Introduction} Hawking's semiclassical
analysis for the black hole  radiation suggests that most of
information in initial states is shield behind the event horizon
and is never back to the asymptotic region far from the
evaporating black hole\cite{HAW1}. This means that the unitarity
is violated by an evaporating black hole. However, this conclusion
has been debated for decades\cite{THOO,SUS,PAG}. The information
loss paradox is closely related to the question of whether the
formation and subsequent evaporation
 of a black hole is unitary. One of the most urgent problems in the black
hole physics is to resolve the unitarity issue.

Recently, Maldacena proposed that the unitarity can be restored if
one takes into account the topological diversity of gravitational
instantons with the same AdS  boundary in (1+2)-dimensional
gravity\cite{MAL}.  Actually, (1+2)-dimensional gravity\cite{BTZ}
is not directly related to the information loss problem because
there is no physical degrees of freedom\cite{CAL}. If this gravity
could be considered to be part of string theory, the AdS/CFT
correspondence\cite{Mal,GKP,Wit} requires that the black hole
formation and evaporating process  be unitary because its boundary
can be described by a unitary CFT. On later, Hawking has withdrawn
his argument on the information loss and suggested  that the
unitarity can be restored  by extending Maldacena's proposal to
(1+3)-dimensional gravity system\cite{HAW2}. The topological
diversity is credited with the restoration of the S-matrix
unitarity in the formation and evaporation
 of a black hole. In this approach of the Euclidean path integral, the pure AdS space
plays an important role in restoring unitarity.  However, the
proposal which is to resolve the information loss paradox by
summing over bulk topologies seems to be failed even in the
(1+2)-dimensional model\cite{KPR,SOL1}.

Since the (1+2)-dimensional gravity and its boundary CFT can
provide a prototype  to compute quasinormal modes and boundary
correlators exactly\cite{BSS1}, they play a crucial role   in
investigating the unitarity issue.   Solodukhin has tried to find
an alternative view to resolving the unitarity problem by
introducing a non-classical deformation of the BTZ black hole
which resembles the geometry of wormhole\cite{SOL2}. As a result
of disappearing the event horizon,  real and discrete modes are
found\cite{SIOP,SOL1}, which means that the considering system
turns out to be unitary.

In this work we focus on the study of the (1+2)-dimensional wave
equation for a scalar field  which include a massless
scalar\cite{BSSS} as well as  a dilaton derived from string
theory\cite{HW,HH,LMM,LKM1,LKM2}.  The pure AdS spacetime provides
a unitary evolution, but it is topologically trivial.  It is
important to find a topologically non-trivial spacetime which
provides a unitary evolution. This work may put a further step to
understand the unitarity  in a semiclassical way because we
propose the  two extremal BTZ black holes (MBTZ, EBTZ) for the
unitarity systems. Our method consists of two steps:  the
potential analysis using the Schr\"odinger-equation and obtaining
 its eigenvalue $E=\omega^2$  by solving the wave
equation exactly. Actually, we translate the black hole-unitary
problem into the boundary-value problem in the Schr\"odinger-like
equation. If the Schr\"odinger operator ${\cal L}=-d^2/dr_*^2
+V(r_*)$ is self-adjoint (${\cal L}^{\dagger}={\cal L}$), its
eigenvalue is real upon imposing  appropriate boundary
conditions\cite{ARFK,CM}.   In this case there is no information
loss and the unitarity is preserved. If one finds  quasinormal
modes, the information is lost during evolution and thus  the
system is not unitary\footnote{We just study the wave propagation
to test how an object (the black hole) responds to an external
perturbation. In the case of  black hole physics, it is impossible
to investigate the interior region of the event horizon using the
Klein-Gordon equation for a scalar.  Hence the radial part of the
Klein-Gordon equation leads to the Schr\"odinger-like equation but
not the exact Schr\"odinger equation because the Klein-Gordon
equation belongs to the relativistic wave equation. The use of
quantum mechanical terminology is here an analogy to understand
the external perturbation intuitively. The system under
consideration is not an exact quantum system. Therefore  we don't
need to do a further work,  the self-adjoint extension of the
Schr\"odinger operator, even if the quasinormal mode is found.}.
In asymptotically AdS spacetime\cite{HHU}, quasinormal modes are
 defined as the solutions which are purely ingoing wave
 at the event horizon
and those which vanish  at infinity because the  potential is
growing at infinity.  The last condition means that any leakage of
the energy (information) is not allowed  through the boundary at
infinity. More precisely, we use the flux boundary condition: the
ingoing flux
 at the horizon and the vanishing  flux
 at infinity\cite{MK}.

In a (1+2)-dimensional spacetime, the Einstein equation with a
negative cosmological constant $\Lambda=-1/\ell^2$ and $8G_3=1$
provides the BTZ solution:
 \beq
 ds_{BTZ}^2=-\Big(-M+
\fr{r^2}{l^2}+\fr{J^2}{4r^2}\Big) dt^2 + \Big(-M+
\fr{r^2}{l^2}+\fr{J^2}{4r^2} \Big)^{-1}dr^2 +r^2
\Big(d\phi-\fr{J}{2r^2}dt\Big)^2,
 \label{1eq2}
 \eeq
where $M$ and $J$ turn out to  be the mass and angular momentum,
respectively\cite{BTZ}. The above metric allows the two horizons
of  $ r^2_{\pm}=Ml^2(1\pm \Delta)/2$ with
$\Delta=(1-J^2/M^2l^2)^{1/2}$. Here the conditions of $|J| \le Ml$
and $M \ge 0$  are required to have the black hole spacetime. Its
thermodynamic quantities  of energy, Hawking temperature,
Bekenstein-Hawking entropy, and heat capacity are given by
$E=M,~~T_{H}=M \Delta/2\pi r_+,~~S_{BH}=4\pi r_+,~~C_J=4\pi
r_+\Delta/(2-\Delta)$~\cite{CLZ,CC,MYU}. We note that the heat
capacity of BTZ black hole is always positive. Hence the BTZ black
hole can be thermally in equilibrium with any size of heat
reservoir. This explains thermodynamically why the BTZ black hole
belongs to an eternal black hole\cite{MAL,MK}. In this work we
consider four interesting cases. i) The non-rotating BTZ black
hole (NBTZ) with $J=0$: $r^2_{+}=l^2M,~T_{H}=\fr{r_+}{2\pi
l^2},~C_{J}=4\pi r_+=S_{BH}$. ii) The pure AdS  spacetime (PADS)
with $M=-1,J=0$: $T_{H}=C_{J}=S_{BH}=0$.  This case corresponds to
the spacetime picture of the NS-NS vacuum state~\cite{MS}. iii)
The massless BTZ black hole (MBTZ) with $M=J=0$:
$T_{H}=C_{J}=S_{BH}=0$. This is called the spacetime picture of
the R-R vacuum state. iv) The extremal BTZ black hole (EBTZ) with
$|J|=lM$: $r^2_{+}=r^2_{-}=l^2M/2,~T_{H}=C_{J}=0,S_{BH}= 4\pi
r_+$.

\section{NBTZ VS PADS} We start with the wave propagation for a
massive scalar field with mass $m$

\beq (\nabla^2 -m^2) \Phi=0 \label{1eq3}
 \eeq
in the background of  the non-rotating BTZ black hole.  Its line
element is given by $ ds_{NBTZ}^2=-(-M+ r^2/l^2) dt^2 + (-M+
r^2/l^2)^{-1}dr^2 +r^2 d\phi^2$\cite{CL}. On the other hand, the
pure AdS spacetime  is defined by $
 ds_{PADS}^2=-(1+r^2/l^2) dt^2 + (1+r^2/l^2)^{-1}dr^2 +r^2
d\phi^2$. Here we set $M=l=1$ for simplicity, unless otherwise
stated. Assuming a mode solution\footnote{There is no a globally
defined time-like Killing vector in the AdS black hole spacetime.
But a time-like Killing vector of $\partial/\partial t$ is future
directed at the region I, the Fig. 3 in Ref.\cite{BTZ}. This means
that an appropriate time evolution is allowed if ${\cal
L}_{\partial t}\Phi=-i\omega \Phi$ with $\omega>0$ in this
work\cite{BD}.} \beq \label{3eq3}
 \Phi(r,t,\phi)=f(r)
e^{-i \omega t} e^{i \ell\phi},~\ell \in {\bf Z} \eeq we find the
radial equation  \beq
 \label{4eq3}
(r^2 \mp1)f''(r) + \Big(3r\mp \fr{1} {r}\Big) f'(r) + \Big[
\fr{\omega^2}{r^2 \mp1} -\fr{\ell^2}{r^2} -m^2 \Big] f(r)=0, \eeq
where the prime ($'$) denotes the derivative with respect to its
argument. The upper (lower) signs denote the NBTZ (PADS) cases.
Introducing $f(r)=\tilde{f}(r)/\sqrt{r}$, the above equation
reduces to \beq \label{5eq3} (r^2 \mp1)\tilde{f}''(r)
+2r\tilde{f}'(r) + \Big[\fr{\omega^2}{r^2
\mp1}-\fr{3}{4}\mp\fr{1}{4r^2} -\fr{\ell^2}{r^2} -m^2 \Big]
\tilde{f}(r)=0 \eeq  which  is suitable for the potential
analysis.

First of all,  it is important to  see how a scalar wave
propagates in the exterior of the NBTZ. For this purpose, we
introduce a tortoise coordinate $ 2r_*= \ln[(r-1)/(r+1)]( r=-\coth
[r_*])$\cite{CM,MK}. We have $r_* \to -\infty (r \to r_+)$ and
 $r_* \to 0 (r \to \infty)$. We transform Eq.(\ref{5eq3})
into the Schr\"odinger-like equation with the Schr\"odinger
operator ${\cal L}_{NBTZ}$ and energy $E=\omega^2$
 \beq\label{7eq3}
  -\fr{d^2}{d r_*^{2}} \tilde{f} + V_{NBTZ}
(r_*)\tilde{f} \equiv {\cal L}_{NBTZ} \tilde{f}=  E \tilde{f}
 \eeq
where  the NBTZ potential is given by\cite{CL}
 \beq V_{NBTZ}(r_*) = \Big[
\Big(\fr{3}{4}+m^2\Big)\coth^2 [r_*] -m^2+\ell^2-\fr{1}{2}-(\ell^2
+\fr{1}{4})\tanh^2 [r_*] \Big]. \label{9eq3} \eeq
 We observe that the potential  decreases exponentially to zero ($V_{NBTZ}\sim e^{-2r_*}$)
  as one approaches the event horizon ($r_* \to
 -\infty)$, while it goes  infinity ($V_{NBTZ} \sim 1/r_*^2$) as one approaches
infinity ($r_* \to 0$). $V_{NBTZ}(r_*)$ looks like the right-half
of  $\cup$. A plane wave appears near the event horizon, whereas a
genuine travelling wave does not appear at infinity.  In order to
obtain the solution which is valid for whole region outside the
black hole, we   solve equation (\ref{4eq3}) directly by
transforming it into a hypergeometric equation. With
$z=(r^2-1)/r^2$, our working region is between $z=0$ and $z=1$,
covering the exterior of the NBTZ. Eq.(\ref{4eq3}) takes a form

\beq z(1-z)f''(z) +(1-z) f'(z) + \fr{1}{4} \Big[ \fr{\omega^2}{z}
-\fr{\ell^2}{1-z} -m^2 \Big] f(z)=0. \label{14eq3} \eeq In order
to obtain  quasinormal modes~\cite{BSSS}, we use  the flux
boundary condition\footnote{Actually, the radial flux of $\Phi$ is
expressed in terms of $f(z)$ as ${\cal F}(z=z_0)\equiv 2\fr{2
\pi}{i} [f^*z\partial_z f-f z\partial_z f^*]|_{z=z_0}$. As in
quantum mechanics, this measures the particle current (flow of
energy or information). Hence this quantity is usually used to
calculate the black hole greybody factor and quasinormal modes.} :
the ingoing flux $({\cal F}_{in}(z=0)<0)$ at the horizon and the
vanishing flux (${\cal F}(z=1)=0$) at infinity. Then we find two
types of quasinormal modes with AdS curvature radius $l$ \beq
\label{27eq3} \omega_{1/2}= \pm \fr{\ell}{l} -i\fr{2}{l}
\Big(n+s_+ \Big)\eeq which means that the operator ${\cal
L}_{NBTZ}$ is not self-adjoint. Here we have $2s_{+}=1+
(1+m^2l^2)^{1/2}$. The discreteness comes from the fact that the
NBTZ is a compact (finite) system. Decomposing $\omega_{1/2}=\pm
\omega_R -i\omega_I$, $w_I$ should be positive because the
corresponding mode decays into the horizon. This bulk perturbation
decays,  as does in the linear response of conformal field
theory\cite{BSS1}. The presence of quasinormal modes is a
mathematically precise formalism of the lack of unitarity in the
semiclassical approach to the bulk system.

To find a unitary system, we study the pure AdS spacetime which
does not contain any topologically distinct object. First  we
transform the wave equation (\ref{5eq3}) with lower signs into the
Schr\"odinger-like equation~\cite{BR1}. We introduce a coordinate
$ r_*=\tan^{-1}[r](r=\tan [r_*])$ to transform Eq.(\ref{5eq3})
into the Schr\"odinger-like equation (\ref{7eq3}) with the energy
$E=\omega^2$ and  PADS potential
 \beq V_{PADS}(r_*) = \Big[
\Big(\ell^2-\fr{1}{4}\Big)\cot^2 [r_*]+
m^2+\ell^2+\fr{1}{2}+\Big(\fr{3}{4} +m^2\Big)\tan^2[r_*] \Big]
\label{9eq4} \eeq which is defined on a box between $r_*=0$ and
$r_*=\pi/2$.
 It is observed that for $m^2>0$,
  $V_{PADS}(r_*)$ increases  to infinity as one approaches $r_* \to
 0(r \to 0$), while it also goes  infinity as one approaches
$r_* \to \pi/2(r \to \infty$). $V_{PADS}(r_*)$ looks like $\cup$.
Using the WKB prescription, we expect to have oscillating modes
between two turning points for $\omega$ which satisfies $\omega^2
> V_{PADS}^{min}=2(m^2+\ell^2)+1$ and $\omega^2=V_{PADS}(r_*)$.

 In order to obtain an explicit form for the
frequency $\omega$, we  solve equation (\ref{4eq3}) with lower
signs directly by transforming it into a hypergeometric equation.
With $z=r^2/(1+r^2)$, our working region is also between $z=0$ and
$z=1$. Eq.(\ref{4eq3}) takes a form

\beq z(1-z)f''(z) +(1-z) f'(z) + \fr{1}{4} \Big[ \omega^2
-\fr{\ell^2}{z} -\fr{m^2}{1-z} \Big] f(z)=0. \label{15eq4} \eeq We
require that the wave function  be zero at infinity because the
potential diverges at infinity.  This  means that the wave
function is normalizable and its flux is zero: ${\cal F}(z=1)=0$.
In addition, requiring a regular solution at $z=0(r=0)$ lead to
real and discrete modes with AdS curvature radius $l$ \beq
\label{27eq4} \omega_{1/2}= \pm \fr{\ell}{l} \pm \fr{2}{l}
\Big(n+s_+ \Big) \eeq which means that ${\cal L}_{PADS}$ is
self-adjoint. The discreteness comes from the finite PADS system.
These normal modes are consistent with those found in the AdS
approach \cite{BKL}. Also this oscillating behavior of a bulk
perturbation is mirrored by the oscillating behavior of the
CFT-boundary approach \cite{BSS2}. It is obvious that we cannot
find any complex mode because there is no the event horizon
(dissipative object).

\section{MBTZ}

 Although
the PADS provides a unitary evolution, it is a topologically
trivial spacetime. It is important to find a topologically
non-trivial spacetime which provides a unitary evolution. One
candidate is the massless BTZ black hole. We start with the wave
 equation (\ref{1eq3})
in the background of  the massless BTZ black hole spacetime: $
ds_{MBTZ}^2=- (r/l)^2dt^2 + (l/r)^2 dr^2 +r^2 d\phi^2$. Assuming a
mode solution (\ref{3eq3}) with $l=1$, the wave equation for
$f(r)$ is given by \beq
 \label{3eq5}
r^2 f''(r) +3r f'(r) + \Big[ \fr{\omega^2-\ell^2}{r^2} -m^2 \Big]
f(r)=0. \eeq Because this equation is so simple, it is not easy to
make a potential analysis. Introducing $r_*=1/r^2$,
Eq.(\ref{3eq5}) can be rewritten as the Schr\"odinger-like
equation (\ref{7eq3}) with the zero energy $E=0$ and the potential
$V_{MBTZ}$  \beq  \label{5eq5} V_{MBTZ}(r_*)=-\fr{k_1}{r_*}
+\fr{k_2}{r_*^2},~k_1=\fr{\omega^2-\ell^2}{4},~k_2=\fr{m^2}{4}.
\eeq It seems that the MBTZ case could not be translated into the
boundary value problem  because its eigenvalue
 is  determined as  $E=0$ initially.
 Near infinity at $r_*=0(r=\infty)$, one finds an approximate
equation of $ d^2f_0(r_*)/dr_*^{2}- (k_2/r_*^2)f_0(r_*)= 0$
 whose solution is given by
\beq f_{0}(r_*) =A_{MBTZ}~ r_*^{s_+} +
B_{MBTZ}~r_*^{s_-}\Big(f_{\infty}(r) =A_{MBTZ}~ r^{-2s_+} +
B_{MBTZ}~r^{-2s_-}\Big) \label{7eq5} \eeq with $s_{\pm}=(1\pm
\sqrt{1+m^2})/2$.
 Here the first is a normalizable mode and the
second is a nonnormalizable mode. However, it is not easy to
obtain approximate solution near the event horizon at
$r_*=\infty(r=0)$ since the potential $V_{MBTZ}$ contains a
long-range interaction term like $k_1/r_*$.

Introducing $z=1/r$, the working region is  extended
 from $z=0$ to $z=\infty$.
 Eq.(\ref{3eq5}) leads to \beq \label{8eq5} z^2 f''(z) -z f'(z)
+ \Big[ (\omega^2-\ell^2)z^2 -m^2 \Big] f(z)=0. \eeq In order to
solve this equation, one first transforms it into the Bessel's
equation with $f(z)=z\tilde{f}(z)$.  Using
$\eta=\sqrt{\omega^2-\ell^2}z$, one finds the Bessel's
equation\beq \label{9eq5} \eta^2 \tilde{f}''(\eta) +\eta
\tilde{f}'(\eta) + \Big[ \eta^2 -\nu^2 \Big] \tilde{f}(z)=0 \eeq
with $\nu=\sqrt{1+m^2}=2s_+-1$. For massless (dilatonic) scalar,
it is given by $\nu=1(3)$. Then we find the waveform which is
valid for whole region between $z=0$ and $z=\infty$\cite{LM} \beq
\label{10eq5}
f(z)=C_1z~J_{\nu}\Big(\sqrt{\omega^2-\ell^2}z\Big)+C_2z
Y_{\nu}\Big(\sqrt{\omega^2-\ell^2}z\Big). \eeq In the limit of
$z\to 0(r \to \infty)$, one has \beq \label{11eq5} f(z) \to
f_0(z)=A_{MBTZ}z^{2s_+}+ B_{MBTZ}z^{2s_-}=A_{MBTZ}r^{-2s_+}+
B_{MBTZ}r^{-2s_-}, \eeq where $f_0(z)$ is consistent with the
approximate solution $f_{\infty}(r)$ in Eq.(\ref{7eq5}). Hence we
choose $B_{MBTZ}=0(C_2=0)$ by imposing the boundary condition at
infinity of $r=\infty$.

Near the event horizon at  $z= \infty(r= 0)$, one has \beqa
\label{12eq5} &&f(z) \to f_{\infty}(z)= C_1 \sqrt{z}
\cos\Big(\sqrt{\omega^2-\ell^2}z-\pi s_+\Big)\\
&&= C_1\fr{\sqrt{z}}{2}\Big[ e^{i\Big(\sqrt{\omega^2-\ell^2}z-\pi
s_+\Big)}+ e^{-i\Big(\sqrt{\omega^2-\ell^2}z-\pi s_+\Big)}
\Big]\equiv f^{in}_{MBTZ}+f^{out}_{MBTZ},\label{13eq5}
 \eeqa where the first term is an ingoing mode and the last is an
 outgoing mode. In this case we have
 $f^{out}_{MBTZ}=[f^{in}_{MBTZ}]^*$ and thus $f_{\infty}(z)$ is
 real. Then it means  that the total flux near the event horizon is zero.
  However, there exist an ingoing flux and an outgoing flux such that
 ${\cal F}^{in}(z=\infty)+{\cal F}^{out}(z=\infty)=0$. Therefore, we cannot obtain the
 wanted case: ${\cal F}^{in}(z=\infty)\not=0,~{\cal F}^{out}(z=\infty)=0$
 because the event horizon is a degenerate point and is located at
 $r=0$.  In other words, there is no room to determine the frequency $\omega$
 since the spectrum of $E$ is set to be zero initially.
This implies that  ${\cal L}_{MBTZ}$ is self-adjoint and the MBTZ
is unitary during evolution. At this stage  we have to distinguish
between the eigenvalue $E$ of ${\cal L}_{MBTZ}$  and the own
frequency $\omega$ for $\Phi$. The continuous frequency reflects
that the MBTZ is an infinite (non-compact) system.
 We conclude
that  its frequency remains  real and continuous\cite{BR2}.
  In order to study
the extremal black hole further, we need to introduce the other
extremal BTZ black hole in the next section.

 \section{EBTZ} We study the
wave propagation for a massive scalar field in the background of
the extremal BTZ black hole given by $ds_{EBTZ}^2=-[(r/l)^2-
2(r_+/l)^2] dt^2 + [r^2l^2/(r^2-r_+^2)^2]dr^2-[2r^2_+/l]dtd\phi
+r^2 d\phi^2$~\cite{CC,KV}. In this case $g^{rr}$ is   also
degenerate at the event horizon of $r=r_{+} =l \sqrt{M/2}$.
Assuming a mode solution in Eq.(\ref{3eq3}), the radial equation
for $f(r)$ is \beqa
 \label{4eq6}
&& \fr{(r^2 -r^2_+)^4}{r^2l^4}f''(r) +
\fr{(r^2 -r^2_+)^3(3r^2+r^2_+)}{r^3l^4} f'(r)  \\
&& +\Big[ \fr{1}{l^2}(\omega l+\ell) \Big((\omega l-\ell)r^2+2\ell
r^2_+ \Big)-\fr{m^2}{l^2}(r^2-r_+^2)^2 \Big] f(r)=0. \nonumber
\eeqa Choosing a good coordinate $z=r^2_+/(r^2-r_+^2)$, the above
equation reduces to the Schr\"odinger-like equation
(\ref{7eq3})\cite{GM,CLS}. Here the potential $V_{EBTZ}$ and its
energy $E=k_0^2$ are given by \beq \label{6eq6}
V_{EBTZ}(z)=-\fr{k_1}{z}
+\fr{k_2}{z^2},~k^2_0=\Omega^2_+,~k_1=\Omega_+\Omega_-,~k_2=\fr{m^2l^2}{4}
\eeq with $\Omega_{\pm}=(\omega l\pm \ell)/\sqrt{2M}$. We comment
that the Schr\"odinger-like equation  for the MBTZ can be obtained
from the EBTZ-equation  by substituting  $z$ and $E=k_0^2$ into
$r_*$ and $E=0$. Thus we include  the previous MBTZ as the special
case of the  EBTZ with  $E=0$.

First we may consider a naively approximate equation of
$d^2f_{\infty}(z)/dz^2+ k^2_0 f_{\infty}(z)=0$ near the horizon at
$z\to \infty(r=r_+)$ whose solution is given by a plane wave \beq
\label{8eq6} f_{\infty}=C_{EBTZ}~e^{i\Omega_{+}
z}+D_{EBTZ}~e^{-i\Omega_{+} z},\eeq where the first term
corresponds to an ingoing mode and the last is an outgoing one. On
the other hand, near infinity at $z\to 0(r=\infty)$ one obtains an
approximate equation $ d^2f_{0}(z)/dz^2-(k_2/z^2) f_{0}(z)=0 $
which gives us   a  solution of $f_{0}(z) =A_{EBTZ}~ z^{s_+} +
B_{EBTZ}~z^{s_-}$.  Here the first term is a normalizable mode and
the second is a nonnormalizable mode.

Up to now we obtain  approximate solutions near $z=\infty, 0$.
However, we don't know whether these are true solutions because of
the long-range potential $V_{EBTZ}$.  In order to obtain the
solution which is valid for whole region outside the EBTZ, we have
to solve equation (\ref{4eq6}) explicitly. Plugging
$f(z)=f_{\infty}(z)f_{0}(z)\tilde{f}(z)$ with
$D_{EBTZ}=B_{EBTZ}=0$ into Eq.(\ref{4eq6}), it takes the form with
$\xi=-2i\Omega_{+}z$
 \beq \xi \tilde{f}''(\xi) +(2s_+ -\xi) \tilde{f}'(\xi) -
\Big(s_+ -\fr{i\Omega_-}{2} \Big)\tilde{f}(\xi)=0. \label{12eq6}
\eeq This corresponds to the confluent hypergeometric equation and
its solution is given by \beq \label{13eq6} \tilde{f}(z)=F[s_+
-\fr{i\Omega_-}{2},2s_+;-2i\Omega_+z]. \eeq Considering the
Kummer's transformation of $F[a,c;\xi]=e^{\xi} F[c-a,c;-\xi]$ with
$a=s_+ -\fr{i\Omega_-}{2}$ and $c=2s_+$, it is easy to show that
the mode solution $f(z)$ is real: $[f(z)]^*=f(z)$.

We choose an  ingoing mode  near $z=\infty$ and a normalizable
solution at $z=0$  as the solution which is valid for whole region
outside the horizon
 \beq
 f(z) \sim e^{i\Omega_{+} z} z^{s_+} F[a,c;-2i\Omega_+z].
 \label{14eq6} \eeq
First we  calculate the flux at $z=0(r=\infty)$. In this case it
confirms that $f_0(z)\sim z^{s_+}$ is a real function because
$F[a,c;-2i\Omega_+z] \to 1,~e^{i\Omega_{+} z}\to 1$ as $z \to 0$.
Then it is obvious that the corresponding flux disappears as \beq
{\cal F}(z=0)= 2\fr{2 \pi}{i} [f^*z\partial_z f-f z\partial_z
f^*]|_{z=0}=0. \label{15eq6} \eeq In order to obtain the flux near
the event horizon at $z=\infty(r=r_+)$, we use simply the reality
condition of $[f(z)]^*=f(z)$. Also we find
 \beq \label{16eq6} {\cal F}(z=\infty)=0.
 \eeq
 It is very curious from Eqs.(\ref{8eq6}) and(\ref{16eq6})
 that even though an ingoing  mode of $f_{\infty} \sim e^{i\Omega_+z}$ is present near the
 horizon, its flux is zero. This implies  that we need a further  investigation on
 the wave propagation in the EBTZ background.  Also it suggests that there is no restriction on
the frequency of $\omega$. Thus its mode  is real and continuous.

Now let us derive an explicit waveform near the event horizon at
$z=\infty(r=r_+)$. For this purpose we introduce the asymptotic
expansion of the confluent hypergeometric function for purely
 imaginary argument $\xi=-2i\Omega_+$ and large $|\xi|$\cite{AS}
\beq \label{17eq6} F[a,c;\xi] \to
\fr{\Gamma(c)}{\Gamma(c-a)}|\xi|^{-a} e^{ \pm i\pi a/2}+
\fr{\Gamma(c)}{\Gamma(a)}|\xi|^{a-c} e^{\pm i\pi (a-c)/2}\eeq
where the upper sign being taken if $-\pi/2 <{\rm arg} (\xi) <
3\pi/2$ and the lower one if $-3\pi/2 <{\rm arg} (\xi) \le
-\pi/2$. Using the above formula,  we can easily prove that the
Kummer's transformation of $F[a,c;\xi]=e^{\xi} F[c-a,c;-\xi]$ is
also valid for large $|\xi|$. The approximate wave function is
given by \beqa \label{18eq6} &&f(z)=e^{i\Omega_{+} z} z^{s_+}
F[a,c;-2i\Omega_+z] \to \\
~~~~&& f_{\infty}(z)=\fr{\Gamma(2s_+)}{\Gamma(s_+
+i\Omega_-/2)}(2\Omega_+)^{-s_+}~e^{-\fr{\pi\Omega_-}{4}}~ e^{
i\Big[\Omega_+z+\fr{\Omega_-}{2}\ln|2\Omega_+z|-\fr{\pi
s_+}{2}\Big]}\nonumber \\
 ~~~~&&+ \fr{\Gamma(2s_+)}{\Gamma(s_+
-i\Omega_-/2)}(2\Omega_+)^{-s_+}~e^{-\fr{\pi\Omega_-}{4}}~ e^{
-i\Big[\Omega_+z+\fr{\Omega_-}{2}\ln|2\Omega_+z|-\fr{\pi
s_+}{2}\Big]}\equiv f^{in}_{EBTZ}+f^{out}_{EBTZ}.\nonumber \eeqa
Comparing the above  with Eq.(\ref{8eq6}) leads to the fact that
the first term corresponds the ingoing mode and the last one is
the outgoing mode. We observe here that the presence of $k_1$-term
in Eq.(\ref{6eq6})(like Coulomb potential) prevents the ingoing
waveform a plane wave in Eq.(\ref{8eq6})\cite{MERC}. Also we note
that the
 contribution from $k_1$-term to the phases is
a logarithmic function of $z$. Even starting with an ugly form of
$f_0(z)\sim z^{s_+}$, a nearly travelling waveform near the event
horizon is developed after transformation. Further it is important
to confirm that the wave function is real
($f_{\infty}(z)=[f_{\infty}(z)]^*$) near the event horizon because
of $f^{out}_{EBTZ}=[f^{in}_{EBTZ}]^*$.

In order to obtain  quasinormal modes, it requires that the wave
function be a purely ingoing mode near the event horizon and
$f(z=0)=0$ at infinity. Here we obtain  a condition of
$s_+-i\Omega_-/2=-n,~n \in {\bf N}$ from $f^{out}_{EBTZ}=0$. Then
we may find the complex and discrete modes with the AdS curvature
radius $l$ and $M=1$ as  \beq \label{19eq6} \omega=-\fr{\ell}{l}
-i\fr{2\sqrt{2}}{l}\Big(n+s_+\Big).\eeq At the first glance there
may exist quasinormal modes for a massive scalar propagation on
the  EBTZ background. However, this condition leads in turn  to
the zero ingoing flux because the flux expression \beq
\label{20eq6} {\cal F}_{in}(z=\infty) \propto
 \fr{\Gamma(2s_+)}{\Gamma(s_+
+i\Omega_-/2)} \fr{\Gamma(2s_+)}{\Gamma(s_+ -i\Omega_-/2)}\eeq
leads to zero exactly when choosing $s_+-i\Omega_-/2=-n$. This
implies that there is no room to accommodate quasinormal modes of
a massive scalar in the background of the EBTZ. Therefore, we show
that there is no restriction on the frequency $\omega$ and thus it
remains real and continuous. This is an enhanced situation when
comparing it with the MBTZ case. A complete analysis is possible
for the EBTZ, because the size of its event horizon is finite and
it is located at $r_+\not=0$ even it corresponds to a degenerate
event horizon. The absence of  quasinormal modes in the EBTZ
 is consistent with the picture of a stable event
horizon with thermodynamic properties $T_{H}=C_{J}=0,S_{BH}=4 \pi
r_+$. This is so because the presence of quasinormal modes implies
that a massive scalar wave is losing its energy continuously into
the extremal event horizon. Here we mention that the absence of
quasinormal modes for the EBTZ is very similar to the case of the
de Sitter cosmological horizon\cite{BMS,MK}.

The Schr\"odinger operator ${\cal L}_{EBTZ}$ is self-adjoint
because its spectrum is real and continuous. Its continuous
spectrum reflects the fact that the EBTZ is an infinite system.
Consequently, the EBTZ is unitary during evolution without loss of
information.

\section{Summary}

We study the wave equation for  a  massive scalar  in
three-dimensional AdS-black hole spacetimes to understand the
unitarity issues in a semiclassical way. Here we introduce four
interesting spacetimes: the non-rotating BTZ black hole (NBTZ),
 pure AdS spacetime (PADS), massless BTZ black
hole (MBTZ), and  extremal BTZ black hole (EBTZ).
 In the
NBTZ case, one finds  quasinormal modes, whereas one finds real
and discrete modes for the PADS case. The presence of quasinormal
modes means that it shows a leakage of information into the event
horizon (dissipative object) and thus it  signals a breakdown of
the unitarity.
 We can easily achieve the unitarity for the PADS.
 This is not a dissipative system  because the perturbations never
disappears completely and always can be restored within the
Poincar\'{e} recurrence time  $t_P$ as in the motion of
oscillation.

On the other hand, we find real and continuous modes for the MBTZ
and EBTZ cases.  These are unitary  systems. The reasons are as
follows. Firstly, the Schr\"odinger operator becomes self-adjoint
upon imposing the Dirichlet condition at infinity, as the same
condition at the origin of radial coordinate in the Coulomb
scattering in quantum mechanics. Secondly,  the corresponding wave
functions are real in whole region outside the event horizon,
especially for near the event horizon and infinity. This means
that there is no leakage of information into the two boundaries:
event horizon and infinity. This confirms from the fact that their
frequencies are real. Thirdly, we find that the radial flux is
identically zero outside the event horizon, even though their wave
functions are non-zero. Actually, we obtain  the ingoing flux as
well as the outgoing flux, but summing over these gives us the
zero flux near the event horizon exactly. This means that there is
no leakage of information into the event horizon. Hence we argue
that the two extremal BTZ black holes are unitary systems. In this
case we cannot obtain discrete spectra because the two belong to
the non-compact system.

Consequently, we propose   two additional systems MBTZ and EBTZ
for the unitarity system.

 A recent work of Hawking does not
explain where the semiclassical analysis of the black hole breaks
down\cite{HAW2}. In the Euclidean path integral approach, the
contribution from the topologically trivial sector (pure AdS
space), which he had previously neglected, is sufficient to
restore the unitarity. However, his arguments are schematic and
thus requires more detailed computations. This proposal seems to
be incorrect even in the (1+2)-dimensional AdS
spacetimes\cite{KPR,SOL1}. In the higher-dimensional AdS
spacetimes, there exists the Hawking-Page transition between the
AdS black hole and pure AdS space. This is a first-order phase
transition. This supports partly that Hawking's arguments is
correct. In the (1+2)-dimensional AdS spacetimes, there exists a
second-order phase transition between NBTZ and MBTZ (not
PADS)\cite{Myung3}. This may explain why we use the MBTZ instead
of the PADS, even both are unitary systems. At this stage,
however, we don't know how the EBTZ plays a role in resolving the
non-unitarity issue of the non-extremal black hole, the NBTZ.

\section*{Acknowledgement} Y. Myung was supported by the Korea Research Foundation Grant
(KRF-2005-013-C00018). H. Lee was in part supported by KOSEF,
Astrophysical Research Center for the Structure and Evolution of
the Cosmos.


\begin{thebibliography}{99}

\bibitem{HAW1} S. W. Hawking,  Phys. Rev. D {\bf 14},
2460 (1976).

\bibitem{THOO} G. 't Hooft,  Nucl. Phys. B {\bf 335}, 138
(1990).

\bibitem{SUS} L. Susskind, hep-th/0204027.

\bibitem{PAG} D. N. Page, hep-th/0409024.

\bibitem{MAL} J. Maldacena, JHEP {\bf 0304}, 021, (2003) 021
[hep-th/0106112].

\bibitem{BTZ} M. Banados, M. Henneaux, C.Teitelboim  and  J. Zanelli,
Phys. Rev. D {\bf 48}, 1506(1993)[gr-qc/9302012].


\bibitem{CAL} S. Carlip, gr-qc/0503022.


\bibitem{Mal} J. Maldacena, Adv. Theor. Math. Phys.  {\bf 2}, 231 (1998)
[Int. J. Theor. Phys.  {\bf 38}, 1113 (1999)] [hep-th/9711200].
\bibitem{GKP}
S. S. Gubser, I. R. Kelebanov and A. M. Polyakov, Phys. Lett. {\bf
B428}, 105 (1998)[hep-th/9802109].
\bibitem{Wit}
E. Witten, Adv. Theor. Math. Phys.  {\bf 2}, 253 (1998)
[hep-th/9802150].


\bibitem{HAW2}
  S.~W.~Hawking,
  hep-th/0507171.

\bibitem{KPR} M. Kleban, M. Porrati  and  R. Rabadan, JHEP {\bf 0410},
030(2004)[hep-th/0407192].

\bibitem{SOL1} S. Solodukhin, Phys. Rev. D {\bf 71}, 064006 (2005)[hep-th/0501053].

\bibitem{BSS1} D. Birmingham, I. Sachs, and S. Solodukhin,
Phys. Rev. Lett.  {\bf 88}, 151301 (2002)[hep-th/0112055].


\bibitem{SOL2} S. Solodukhin, hep-th/0406130. 


\bibitem{SIOP} G. Siopsis, Class. Quant. Grav. {\bf 22}, 1425 (2005)
[hep-th/0408091].



\bibitem{BSSS} D. Birmingham, I. Sachs, and S. Sen, Phys. Lett.  {\bf B413},
281 (1997)[hep-th/9707188].



\bibitem{HW}  G. T. Horowitz  and  D. L. Welch, Phys. Rev. Lett. {\bf 71},
328(1993)[hep-th/9302126].
\bibitem{HH} J. H. Horne and G.T.
Horowitz, Nucl. Phys. B {\bf 368},
444(1992)[hep-th/9108001]. 



\bibitem{LMM} H. W. Lee and Y.S. Myung, Phys. Rev. D {\bf 58}, 104013(1998)
[hep-th/9804095].
\bibitem{LKM1} H. W. Lee, N. J. Kim, and  Y. S. Myung, Phys. Rev. {\bf D58},
084022(1998)[hep-th/9803080]. 
\bibitem{LKM2}H. W. Lee, N. J. Kim, and  Y. S. Myung,
 Phys. Lett. {\bf B441}, 83(1998)[hep-th/9803227].

\bibitem{ARFK} J. B. Arfken and H. J. Weber, {\it Mathematical
Methods for Physics} (Academic Press, San Diago, 1995), chap.9.


\bibitem{CM} J. S. F. Chan and R. B. Mann, Phys. Rev. {\bf D55},
7546(1997)[gr-qc/9612026].

\bibitem{HHU}G. T. Horowitz and V. Hubeny, Phys. Rev.  {\bf D62},
024027 (2000)[hep-th/9909056].


\bibitem{MK} Y. S. Myung and N. J. Kim, Class. Quant. Grav. {\bf 21},
63(2004)[hep-th/0304231].



\bibitem{CLZ} R. G.  Cai, Z.-J. Lu, Y.-Z. Zhang, Phys. Rev. D {\bf
55}, 853(1997)[gr-qc/9702032].
\bibitem{CC}  R. -G. Cai and  J.-H. Cho,
Phys. Rev. D {\bf 60}, 067502(1999)[hep-th/9803261].
\bibitem{MYU} Y. S. Myung, Phys. Lett. B {\bf 579}, 205(2004)
[hep-th/0310176].

\bibitem{MS}  J. Maldacena and  A. Strominger, JHEP {\bf 9812},
005(1998)[hep-th/9804085].


\bibitem{CL}V. Cardoso and J. P. Lemos, Phys. Rev.  {\bf D63},
124015 (2001)[gr-qc/0101052].

\bibitem{BD} N. D. Birrell and P. C. W. Davies,
 {\it Quantun fields in curved space} (Cambridge Univ., New York,
 1982).

\bibitem{AS} M. Abramowitz and I. Stegun, {\it Handbook of
Mathematical Functions} (Dover Publication, New York, 1970)


\bibitem{BR1}  J. Barbon and  E. Rabinovici,
Fortsch. Phys. {\bf 52}, 642(2004)[hep-th/0403268];
hep-th/0503144.



\bibitem{BKL} V.  Balasubramanian, P. Kraus and A. Lawrence,
 Phys.Rev. D{\bf 59}, 046003 (1999)[hep-th/9805171].

\bibitem{BSS2}  D. Birmingham, I. Sachs, and  S. N. Solodukhin,
Phys. Rev. D{\bf 67}, 104026(2003)[hep-th/0212308].


\bibitem{LO} G. Lifschytz and  M. Ortiz, Phys. Rev. D {\bf 49}, 1929
(1994)[gr-qc/9310008].

\bibitem{LM} H. W. Lee and Y. S. Myung, hep-th/9808002.

\bibitem{BR2}  J. Barbon and  E. Rabinovici,  JHEP {\bf 0311},
047(2003)[hep-th/0308063];


\bibitem{KV} E. Keski-Vakkuri, Phys. Rev. D  {\bf 59}, 104001(1999)
[hep-th/9808037].

\bibitem{GM}  J. Gamboa and F. Mendez, Class. Quant. Grav. {\bf 18},
225(2001)[hep-th/0006020].

\bibitem{CLS}  J. Crisostomo, S. Lepe, and  J. Saavedra, Class. Quant. Grav. {\bf 21},
 2801(2004) [hep-th/0402048].

\bibitem{MERC} E. Merzbacher, {\it Quantum mechanics} (John  Wiley and
Sons, New York, 1961), p.248.

\bibitem{BMS} R. Bousso, A. Maloney, and A. Strominger,
Phys. Rev. {\bf D65}, 104039 (2002)[hep-th/0112218].

\bibitem{Myung3} Y.~S.~Myung,
  Phys.\ Lett.\ B {\bf 624}, 297 (2005)
  [hep-th/0506096].



\end{thebibliography}
\end{document}